\documentclass[12pt,a4paper]{amsart}

\usepackage{graphicx}
\usepackage[centering]{geometry}
\usepackage[round]{natbib}
\newtheorem*{theorem}{Theorem}

\begin{document}

\title[Correcting the apparent rate acceleration under a Jukes-Cantor model]%
{Correcting the apparent mutation rate acceleration at shorter
  time scales under a Jukes-Cantor model}
\author[Tuffley, White, Hendy and Penny]{Christopher
  Tuffley, W Timothy J White, Michael D Hendy and David Penny}
\date{11th April 2012}

\address{Institute of Fundamental Sciences, Massey University,
         Private Bag 11222, Palmerston North 4442, New Zealand}
\email{c.tuffley@massey.ac.nz}

\address{Department of Mathematics and Statistics, University of Otago, 
         P.O.\ Box 56, Dunedin 9054, New Zealand}

\address{Institute of Molecular Biosciences, Massey University,
         Private Bag 11222, Palmerston North 4442, New Zealand}

\address{Department of Mathematics and Statistics, University of Otago, 
         P.O.\ Box 56, Dunedin 9054, New Zealand}

\begin{abstract}
At macroevolutionary time scales, and for a constant mutation rate,
there is an expected linear relationship between time and the number
of inferred neutral mutations (the ``molecular clock'').
However, at shorter time scales a number of recent studies have observed
an apparent acceleration in the rate of molecular evolution. 

We study this apparent acceleration under a Jukes-Cantor model 
applied to a randomly mating population, and show that, under the
model, it arises as a consequence of ignoring short term effects due
to existing diversity within the population. The acceleration can be
accounted for by adding the correction term $h_0e^{-4\mu t/3}$ to
the usual Jukes-Cantor formula $p(t)=\frac34(1-e^{-4\mu t/3})$, where
$h_0$ is the expected heterozygosity in the population at time $t=0$. 
The true mutation rate $\mu$ may then be recovered, even if $h_0$
is not known, by estimating $\mu$ and $h_0$ simultaneously using
least squares. 

Rate estimates made without the correction term (that is, incorrectly
assuming the population to be homogeneous) will result in a divergent
rate curve of the form $\mu_\mathrm{div}=\mu+C/t$, so that the
mutation rate appears to approach infinity as the time scale
approaches zero.  While our quantitative results apply only to the
Jukes-Cantor model, it is reasonable to suppose that the qualitative
picture that emerges also applies to more complex models. Our study
therefore demonstrates the importance of properly accounting for any
ancestral diversity, as it may otherwise play a dominant role in rate
overestimation.
\end{abstract}

\maketitle 

\section{Introduction}

It is well-known since Kimura~\citeyearpar{Kimura1968,Kimura1983} that over
time scales of tens and hundreds of millions of years the longer term
rate of molecular evolution, $k$, is expected to be virtually the same
as the short term rate of neutral mutations $\mu$, that is,
$k\approx\mu$.  However, at shorter time scales a number of recent
studies
(for example~\citet{Garcia-Moreno2004,Ho2005,Penny2005,Millar2008,Henn2009};
although there were some earlier indications~\citep{Fitch1985})
have observed an apparent acceleration in the rate of molecular
evolution. This has led to debate as to the underlying causes
of the apparent acceleration, and raised important questions 
as to how long it persists and how to correct for it.

A number of factors that may contribute to this apparent rate
acceleration have been proposed, and we refer the reader to Ho et
al.~\citeyearpar{Ho2011} for a recent review. One such factor is the
effect of ancestral polymorphism \citep{Peterson2009,Charlesworth2010}. Indeed, it is clear that, if
ancestral polymorphism is present and not properly accounted for, then
an apparent rate acceleration will inevitably be seen.  Consider a
comparison between two sequences drawn from a population at time $t=0$
which is incorrectly assumed to be homogeneous. Any differences
between the sequences due to polymorphism will appear to be mutations
that have occurred in zero time, leading to an apparent infinite
substitution rate at time zero.  Moreover, on sufficiently short time
scales the average difference due to polymorphism will dominate any
changes due to substitution, and to first order the expected
difference between sequences at times $0$ and $t$ will be given by
$p(t)=C+\mu't$. (Here $\mu'$ will differ slightly from the mutation
rate $\mu$, as some mutations will act to decrease polymorphism.)
Naively dividing by $t$ to recover $\mu$, under the assumption that
the first order approximation is given by $p(t)=\mu t$, then gives an
apparent divergent rate curve $\mu'+C/t$, suggesting that unaccounted
ancestral polymorphism will contribute a term of the form $C/t$ to
rate estimates. This \emph{a priori} estimate is in sharp contrast
with the use of exponential decay curves of the form $k+\mu
e^{-\lambda t}$ (for example, by Ho~et~al.~\citeyearpar{Ho2005}) to fit
estimated rates.

The effect of ancestral polymorphism on sequence divergence has been
studied quantitatively by Peterson and Masel~\citeyearpar{Peterson2009}, who
calculate the expected divergence between two sequences as a function
of time under the Moran model with selection. Their model provides a
reasonable fit to data from Genner et al.~\citeyearpar{genner2007} and Henn
et al.~\citeyearpar{Henn2009}, confirming that the observed rate acceleration
is consistent with an underlying constant mutation rate.  However,
their model does not admit a simple closed form analytic expression,
and this makes it difficult to correct for the apparent rate
acceleration or quantify its duration.

The purpose of this paper is to quantify the effect of ancestral
polymorphism under a Jukes-Cantor type model applied to a randomly
mating population. This is a more restrictive setting than that
considered by Peterson and Masel, but the benefit of working in this
simpler setting is that we are able to obtain simple and explicit
analytic expressions for quantities of interest. In particular, we
show that, under the model, ancestral polymorphism contributes a term
$h_0e^{-4\mu t/3}$ to the usual Jukes-Cantor formula
$p(t)=\frac34(1-e^{-4\mu t/3})$, where $h_0$ is the expected
heterozygosity at time $t=0$, and confirm that comparisons between
sequences made under the incorrect assumption $h_0=0$ lead to a
divergent rate curve of the form $\mu_\mathrm{div}=\mu+C/t$.  This has
consequences at long time scales as well as short, as $C/t$ tends to
zero comparatively slowly as $t$ tends to infinity (in particular,
more slowly than any exponential decay), and so the effects of
ancestral polymorphism may be relatively long lived. We show however
that the true mutation rate $\mu$ may still be recovered from several
observations even if the level of heterozygosity at time 0 is unknown,
by estimating $h_0$ simultaneously with $\mu$ using least squares. We
show further that our correction term also applies to other population
structures (for example, island models) under an appropriate
assumption on $h_0$.

These results show that ancestral polymorphism, where present and
unaccounted for, will be a significant contributing factor to mutation
rate overestimation, with the magnitude of the effect approaching
infinity as the time scale shrinks to zero.  
Our principal finding then
is that an apparent rate acceleration at short time scales is
consistent with --- \emph{and indeed to be expected under} --- a
constant mutation rate, in the presence of ancestral polymorphism that
is not properly taken into account. This finding is in agreement with
that of Peterson and Masel~\citeyearpar{Peterson2009}. 

Ancestral polymorphism is just one of several processes that are
thought to contribute to the apparent rate acceleration, and is
unlikely to be the sole contributing factor.  However, it is clearly
vital that its effect be quantified and accounted for if we are to
obtain meaningful and accurate rate estimates on short time scales,
and fully resolve the question of the apparent acceleration.

\section{The model, and our main result}

We consider a population of $N$ individuals evolving under a random
mating process, with discrete generations, where at each generation
the allele of each individual is replaced with a copy of the allele
from the previous generation, chosen uniformly at
random~\citep{Wright1931}. For simplicity we restrict our attention to
haploid populations.  We assume that alleles are $r$-state characters
evolving under the $r$-state Jukes-Cantor
model~\citep{Jukes1969}, with an instantaneous mutation rate of
$\mu$ mutations per individual per generation per site.  All mutations
are assumed to be neutral.  We orient time in the forwards direction,
so that populations at times $t>0$ descend from the population at
$t=0$.  We refer to the population at time $0$ as the reference
population, and the population at a given time $t>0$ of interest as
the contemporary population.

Consider a pair of individuals chosen uniformly at random, one from
each of the reference and contemporary populations, and let $P(t)$ be
the probability that they have different character states at a fixed
site.  Suppose that the distribution of character states at that site
in the reference population is given by $\pi=(\pi_1,\ldots,\pi_r)$.
(Note that $\pi$ as used here is the distribution of states within the
population at a given site, rather than the equilibrium distribution
of the Jukes-Cantor model, which is the distribution of states across
all sites within an individual.)  Then

\begin{theorem}
The probability that uniformly randomly chosen members of the
reference and contemporary populations have different states
at a given site is given by
\begin{equation}
\label{Psitesdiffer.eq}
P(t) = h_0e^{-\frac{r}{r-1}\mu t} + \tfrac{r-1}{r}(1-e^{-\frac{r}{r-1}\mu t}),
\end{equation}
where
\[
h_0 = P(0) = \sum_{i=1}^r \pi_i(1-\pi_i)
    = \sum_i\sum_{j\neq i} \pi_i\pi_j
\]
is the expected heterozygosity at $t=0$. In particular, for $r=4$ (as 
for nucleotides) 
we get
\[
P(t) = h_0e^{-\frac{4}{3}\mu t} + \tfrac{3}{4}(1-e^{-\frac{4}{3}\mu t}).
\]
\end{theorem}

The proof of the theorem is given in the Appendix.  Note that the
second term in equation~\eqref{Psitesdiffer.eq} is the standard
probability under the $r$-state Jukes-Cantor model that a net change
takes place at the given site, when the contemporary sequence directly
descends from the reference sequence --- this second term is the usual
form of the equation for longer time periods. Thus the theorem adds
the correction term $h_0e^{-r\mu t/(r-1)}$ when the sequences being
compared are not necessarily direct descendants. This accounts for the
variation that is present in the reference population from past
mutations that have not yet been either fixed or lost. The two
expressions $P(t)$ and the Jukes-Cantor mutation probability are
asymptotic, in the sense that 
\[
\frac{P(t)}{\tfrac{r-1}{r}(1-e^{-\frac{r}{r-1}\mu t})}
= 1 + \frac{h_0e^{-\frac{r}{r-1}\mu t}}
            {\tfrac{r-1}{r}(1-e^{-\frac{r}{r-1}\mu t})} \rightarrow 1
\]
as $t\to\infty$.

Note that the theorem requires that the population at the earlier time
be used as the reference population. In particular, if populations at
several different times are to be compared, then this should be done
by comparing each population against the oldest population, to ensure
that all pairwise comparisons made involve the same value of the
initial  heterozygosity $h_0$.
This requirement may be relaxed if there is reason to believe that the
heterozygosity $h$ is constant, or nearly so. In general, we expect
$\pi$ (and hence $h$) to depend on $t$, but as $t\rightarrow\infty$ we
also expect $\pi$ to approach an equilibrium distribution
$\pi_\infty$, where the rate at which new mutations are introduced
balances the rate at which mutations are fixed or lost by the mating
process. If this equilibrium is assumed to have occurred then
equation~\eqref{Psitesdiffer.eq} may be written in the form
\begin{equation}
\label{P@h_inf.eq}
P(t) = h_\infty e^{-\frac{r}{r-1}\mu t} +
               \tfrac{r-1}{r}(1-e^{-\frac{r}{r-1}\mu t}),
\end{equation}
where $h_\infty$ is the expected heterozygosity at equilibrium. Under
this assumption any two populations at times $t_1$ and $t_2$ may
be compared, using $t=|t_1-t_2|$. 

\subsection{Recovering the mutation rate from observed data}

We now demonstrate how the mutation rate $\mu$ may be
estimated from observations of $P(t)$. If the expected heterozygosity
is assumed to have reached equilibrium, then $h_\infty$ may be
substituted for $h_0$ throughout. 

Equation~\eqref{Psitesdiffer.eq} may be rearranged to the form
\begin{equation}
\label{Pgathered.eq}
P(t) 
= \tfrac{r-1}{r}\left(1-(1-\tfrac{r}{r-1}h_0)e^{-\frac{r}{r-1}\mu t}\right).
\end{equation}
If $h_0$ is known then we may estimate $\mu$ as
\begin{equation}
 \mu =  -\frac{r-1}{rt}\log\frac{1-\tfrac{r}{r-1}P(t)}{1-\tfrac{r}{r-1}h_0}.       
\label{estimateknowingh0.eq}
\end{equation}
On the other hand, if $h_0$ is not known then it will need to be
estimated simultaneously with $\mu$. This may be done using a least
squares fit to 
\[
\log\left(1 - \tfrac{r}{r-1}P(t)\right) 
        = \log(1 - \tfrac{r}{r-1}h_0) - \tfrac{r}{r-1}\mu t;
\]
if the least squares line is $y=mt+b$ then we recover $\mu$ and $h_0$ as
\begin{align*}
\mu &= -\tfrac{r-1}{r}m, & h_0 &= \tfrac{r-1}{r}(1-e^b).
\end{align*}

\subsection{The divergent rate curve}

A divergent rate curve arises when we either omit or use an incorrect value 
for $h_0$
to estimate $\mu$.  In particular, if we ignore the existing
diversity and thus use $h_0=0$ we get the
estimate
\begin{align}
 \mu_\mathrm{div} 
 &= -\frac{(r-1)\log(1-\tfrac{r}{r-1}P(t))}{rt} \label{lazyj.eq} \\ 
 &= \mu - \frac{(r-1)\log\left(1-\tfrac{r}{r-1}h_0\right)}{rt}, \nonumber
\end{align}
which adds the divergent term
$\frac{(1-r)\log\left(1-\tfrac{r}{r-1}h_0\right)}{rt}$ to the estimate
above.  Thus, if the variation present within the reference population
is ignored we obtain a divergent rate curve of the form 
$\mu_{\mathrm{div}}=\mu+C/t$, where 
$C=\frac{1-r}{r}\log\left(1-\tfrac{r}{r-1}h_0\right)$ is a positive
constant independent of $t$. This rate estimate tends
to infinity as $t\to0$, and thus becomes
increasingly inaccurate on shorter and shorter times scales.

\section{Results of simulations}

Figure~1A shows results of simulations over
microevolutionary time scales, and exhibits the phenomena described
above. Most importantly, if we do not correct for existing genetic
diversity there is the apparent acceleration at shorter times, even
though the basic neutral mutation rate $\mu$ is kept constant. This
first main point then is that the apparent increase in ``rate'' (the
divergent rate curve) is obtained at shorter periods. However, it is
then important that, using equation~\eqref{estimateknowingh0.eq} and the
value of $h_0$, the value of $\mu$ can still be estimated accurately
at these shorter intervals. This is certainly as expected; it has
always been assumed that the mutation rate $\mu$ was basically
constant, that it was independent of time. Thus the conclusion from
Figure~1A is that by explicitly considering genetic
variability in the ancestral (reference) population --- either by using
an \emph{a priori} estimate of it (blue curve), or by estimating it
simultaneously using least squares (dashed black line) --- it is
possible to recover the mutation rate $\mu$. In practice, it may be that
knowledge of $\mu$ (from longer term studies) may also be important in
understanding population structure and its change over time.

\subsection{The simulations}

The simulations were run using the statistical package
R~\citeyearpar{R-manual2011}.  Each simulation begins with a haploid
population containing 70 individuals having allele A and 30
individuals having allele B, giving a true $h_0$ of 0.42.  There are
four allele types in total, and mutations from any given allele to any
of the other three take place at equal rates, as in the Jukes-Cantor
model for DNA substitutions.  This population is then evolved for 1000
generations.  In each generation, reproduction is simulated according
to the Wright-Fisher process, whereby each of the 100 individuals
making up the new population is chosen randomly with replacement from
the previous generation; each of these new individuals is then
subjected to mutation at the rate of $\mu = 0.001$ mutations per
individual per generation.  For the final step in each generation, a
pair of individuals --- one from the current population and one from
the initial population --- is picked randomly and compared; if they
have different alleles, a counter for that generation is increased by
1.  The entire simulation was repeated 2000 times, with counters
accumulating across runs: consequently the final counter value for
generation $i$, after dividing by 2000, is an estimate of the
probability that an individual picked randomly from the initial
population differs from an individual picked randomly from generation
$i$.  These relative frequencies are used to estimate the mutation
rate parameter $\mu$ in several different ways, as we now describe.

\subsection{Estimating $\mu$}

Three different approaches to estimating $\mu$ are shown in Figure 1A.
The dashed pink curve shows the result of estimating $\mu$ while
incorrectly assuming $h_0 = 0$ (the assumption made by most previous
studies); the solid blue curve shows the result of estimating $\mu$
assuming $h_0$ to be its true value, 0.42.  The difference is plain,
particularly at early times when the probability of observing a
difference is dominated by the heterozygosity of the initial
population.  To demonstrate the limited accuracy achievable using the
incorrect assumption that $h_0 = 0$, the $\mu$ curve that would be
estimated from an infinite number of observations (instead of 2000) is
shown as a dashed green line --- it is scarcely better, indicating
that sampling error is not the problem.  The fact that the pink curve
tracks the dashed green curve also indicates that, as expected, the
simulation fits our theoretical divergent curve $\mu_\mathrm{div}$ of
equation~\eqref{lazyj.eq}.  Note that the erratic behaviour of the blue
curve near 0 is due to numerical instability, as the calculations
there involve logs and ratios of numbers close to zero.

Both of the above estimation procedures can estimate $\mu$ using a
probability estimate from a single time point, but assume $h_0$ to be
known \emph{a priori}.  Where this is not the case, $h_0$ can be
estimated simultaneously with $\mu$ using least squares, which
requires probability estimates from multiple time points.  The result
of estimating a single value for $\mu$ using all 1001 time points is
shown as a horizontal dashed black line in Figure~1A.

\begin{figure}
\begin{center}
\includegraphics[scale=0.64]{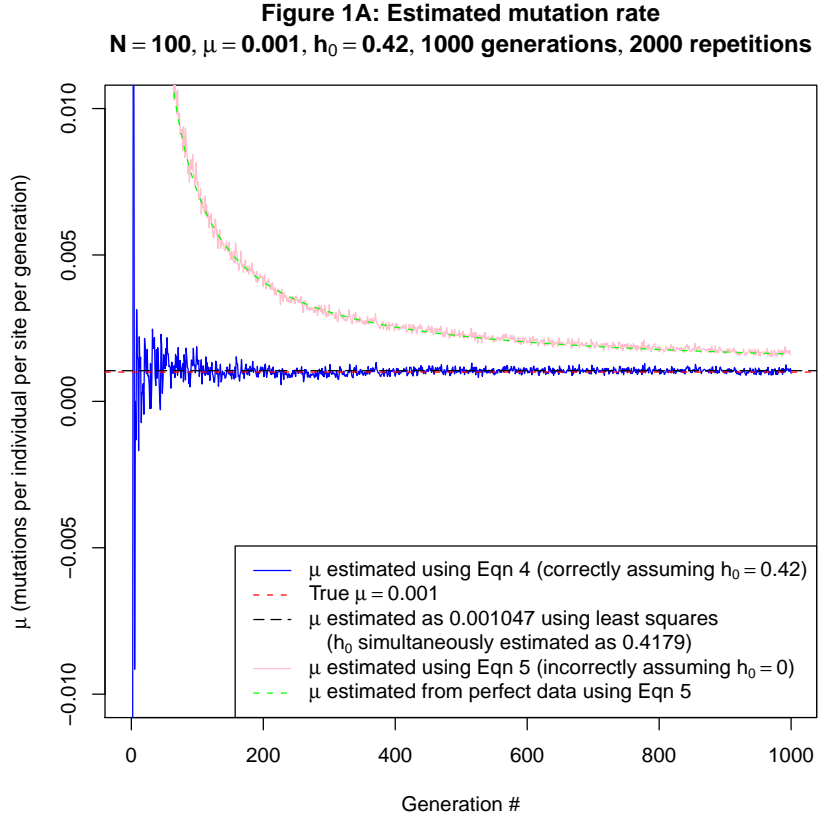}
\caption[Results of simulations]{Estimated mutation rates against time
  for three different population structures.  (1A): A single
  population initially containing 70 individuals having allele A and
  30 individuals having allele B; (1B): A 2-island model, both of
  whose subpopulations are initially the same as for (1A), with 1\%
  migration per generation from island 1 to island 2, and 10\% migration 
per generation in the
  reverse direction; (1C): A 2-island model, whose first subpopulation
  is the same as for (1A), and whose second subpopulation contains 25
  individuals of each of the four states. The graphs show the results
  of simulations of four state data.
\begin{minipage}{14cm}
\begin{description}
\setlength{\rightmargin}{10cm}
\item[Dashed red line] 
the value of the mutation rate $\mu$ used. 
\item[Blue curve] 
$\mu$ estimated at each time point from simulated data transformed
  using equation~\eqref{estimateknowingh0.eq} and the correct value for
  the ancestral heterozygosity $h_0$.
\item[Pink curve] simulated data transformed using the
  incorrect transform (equation~\eqref{lazyj.eq}) to estimate $\mu$, giving
  a divergent rate curve. 
\item[Dashed green curve (1A, 1B)]
  theoretical divergent rate curve, obtained by transforming perfect
  data (equation~\eqref{Psitesdiffer.eq}) according to
  equation~\eqref{lazyj.eq}. This assumes an incorrect value of $h_0$= 0.
\item[Dashed black line] the least squares estimate of $\mu$ from the
  simulated data. This uses no knowledge of $h_0$ or $\mu$ (unlike the
  blue curve).
\end{description}
\end{minipage}
Note that the blue curve behaves poorly near 0, where the calculations
are numerically unstable as they involve logs and ratios of numbers
close to zero.}
\label{graph.fig}
\end{center}
\end{figure}

\begin{figure}
\begin{center}
\includegraphics[scale=0.64]{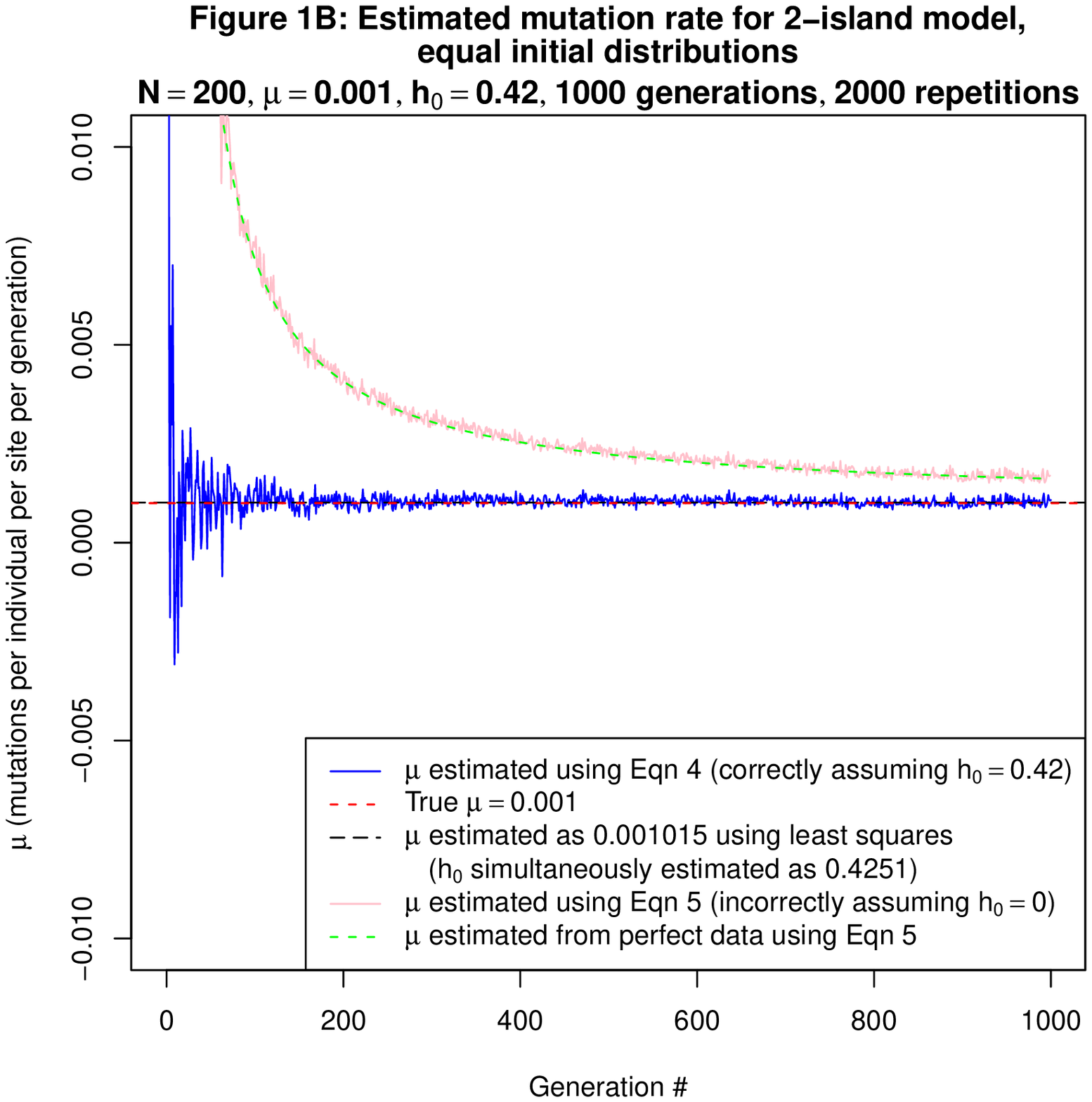}
\end{center}
\end{figure}

\begin{figure}
\begin{center}
\includegraphics[scale=0.64]{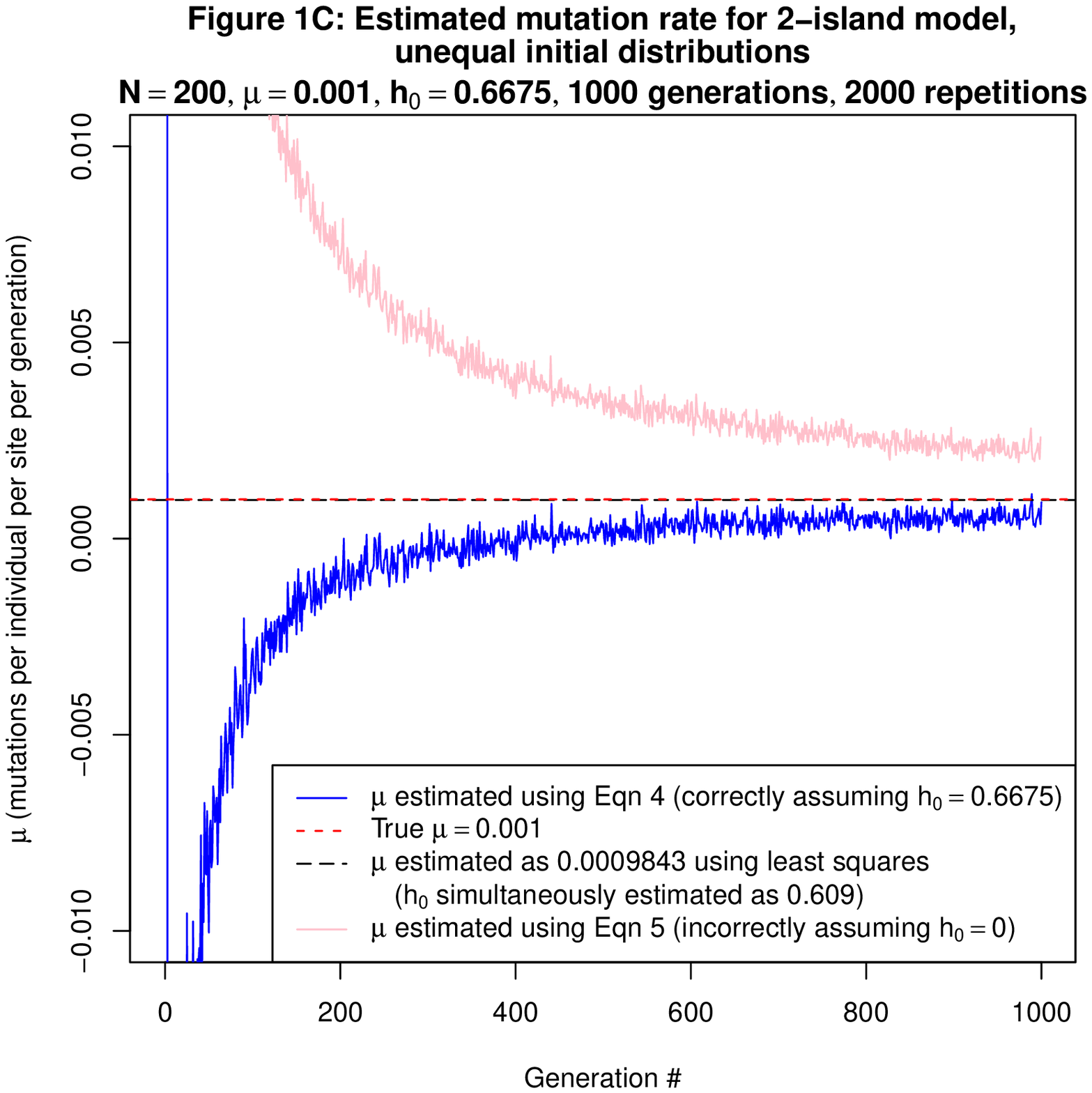}
\end{center}
\end{figure}

\subsection{Other population structures}

Our corrected formula extends to cases where the population consists
of multiple subpopulations that interact via arbitrary migration
rates, provided that the initial distribution of states is the same
for each subpopulation.  Figure~1B shows the accurate
recovery of $\mu$ and $h_0$ for a simulated population consisting of
two islands, each containing 70 individuals having allele A and 30
individuals having allele B, that mix sporadically and asymmetrically:
in each generation, the probability that an individual in
subpopulation 1 comes from subpopulation 2 is just 1\%, while the
probability that an individual in subpopulation 2 comes from
subpopulation 1 is 10\%. Adding this limited kind of population
structure does not alter the fact that the probability distribution of
states for the ancestor of an individual chosen randomly at time $t$
remains equal to that for an individual chosen randomly from the
initial population, so the probability $p_\mathrm{diff}$ that the
ancestral and reference states differ is still given by $h_0$. 

If the initial state distributions differ across subpopulations, then
the probability that the ancestor of a randomly-chosen contemporary
individual has a particular state is no longer constant, so $h_0$ in
equation~\eqref{Psitesdiffer.eq} must be replaced with a function of time,
$p_\mathrm{diff}(t)$.
Figure~1C shows results from a simulation of
one such scenario.  There are two islands, each containing 100
individuals: the first contains 70 individuals having allele A and 30
individuals having allele B, the second contains 25 individuals with
each of the four alleles.  Migration rates are as for
Figure~1B.  The blue curve, which shows an attempt to
estimate $\mu$ using equation~\eqref{estimateknowingh0.eq} and assuming
that $h_0 = 0.6675$ (the probability that two individuals chosen
randomly from the initial population have different states), produces
a divergent rate curve because this more general population structure
violates the assumption that $p_\mathrm{diff}$ is constant and equal
to $h_0$. 

Despite the fact that $p_\mathrm{diff}$ is nonconstant when the
subpopulations have different initial state distributions, our
approach is still useful here. Under reasonable conditions on
migration probabilities, the probability that the ancestor of a
randomly-chosen individual has a given state converges towards an
equilibrium value as $t\to\infty$, and so $p_\mathrm{diff}$ will
approach an equilibrium value also.  In particular, when estimating
$\mu$ and $h_0$ simultaneously via least squares, only the estimate of
$h_0$ is affected by different initial state distributions across
subpopulations: as expected, the dashed black line depicting the least
squares estimate of $\mu$ in Figure~1C is close to the true value of
$\mu$. Since $\mu$ is usually the parameter of interest, this means
that our model can still be used for inference with these more general
population structures.  Note that any edge-weighted graph describing
mating probabilities within a population can be represented as a
multi-island model in which each island contains a single individual.

\section{Conclusions}

The Jukes-Cantor model is one of the oldest and
simplest substitution models, and the benefit of studying 
ancestral polymorphism in this
simple setting is that we are able to obtain explicit analytic
expressions for quantities of interest, including the expected
divergence between two sequences, and the uncorrected rate curve
$\mu_{\mathrm{div}}$. Moreover, many more complex and realistic
models include the Jukes-Cantor model as a special case, so while
our precise quantitative findings are only supported under the 
limited model considered here, there are clear qualitative implications
for these more general models. 

Given the rapid advance in DNA sequencing technology we expect a large
increase in sequences that have diverged in recent, intermediate and
longer times. It is therefore essential to be able to generalise
results to a full range of divergence times --- it is no longer
appropriate to consider separately shorter term microevolutionary and
longer term macroevolutionary studies. Our results show that 
a significant contributing factor 
to the apparent acceleration at shorter times is the pre-existing
diversity within a population, and this can be estimated for a wide
range of population sizes and
structures~\citep{Charlesworth-Charlesworth2010}. By considering the
expected diversity in a population we show how the real rate of
neutral mutation can be estimated at shorter times, using either an
appropriate estimate of genetic diversity or data from multiple time
points.  As the timescales of traditional phylogenetics and population
genetics draw closer together, we anticipate seeing an increasing
emphasis on the careful handling of genetic diversity in phylogenetic
analyses; the work presented here can provide a starting point for the
exploration of more general models involving more complex substitution
processes, time-varying population sizes, and other effects.

\appendix

\section{Proof of the main result}

The individual chosen from the contemporary
population descends from a unique
member of the reference population, and this ancestor is equally
likely to be any member of the reference population. We may therefore
calculate the probability that the contemporary and reference states
differ by considering two cases: the ancestral state agrees with the
reference state, but evolves to disagree; and the ancestral state
disagrees with the reference state, and evolves so as to remain in
disagreement at time $t$. 

Under the $r$-state Jukes-Cantor model the probability that there is
a net transition from state $i$ to state $j\neq i$ in time $t$ is given by
\[
p(t) = \tfrac{1}{r}(1-e^{-\frac{r}{r-1}\mu t}). 
\]
The probability that the ancestral state agrees with the reference state
but evolves to disagree is therefore given by
\begin{equation}
\label{initiallyagree.eq}
\sum_{i} \pi_i^2(r-1)p(t) 
 = (r-1)p(t)\sum_{i} \pi_i^2,
\end{equation}
while the probability that both the ancestral and contemporary states
differ from the reference is given by
\begin{align}
\sum_i\sum_{j\neq i} \pi_i\pi_j(1-p(t)) &= 
(1-p(t))h_0.
    \label{initiallydisagree.eq}
\end{align}
Observe that 
\[
1=\left(\sum_{i}\pi_i\right)^2 
 = \sum_i\pi_i^2+\sum_i\sum_{j\neq i} \pi_i\pi_j = h_0+\sum_i\pi_i^2,
\]
so $\sum_i \pi_i^2 = 1- h_0$. Hence
summing equations~\eqref{initiallyagree.eq} and~\eqref{initiallydisagree.eq}
we get
\begin{align*}
P(t) &= (1-h_0)(r-1)p(t)+h_0\left(1-p(t)\right) \\
     &= (r-1)p(t)+h_0(1-p(t)-(r-1)p(t))   \\
     &= (r-1)p(t)+h_0(1-rp(t)) \\
     &= \frac{r-1}{r}(1-e^{-\frac{r}{r-1}\mu t})
            +h_0e^{-\frac{r}{r-1}\mu t},
\end{align*}
as claimed.

\bibliographystyle{plainnat}
\bibliography{mbe}

\end{document}